\documentclass[pra,aps, twocolumn,showpacs]{revtex4}
\usepackage{}

\usepackage{graphicx}  
\usepackage{epstopdf}
\usepackage{dcolumn}   
\usepackage{bm}        
\usepackage{amssymb}   
\usepackage{amsmath}   
\usepackage{amsthm}
\usepackage{chemarrow}
\usepackage{color}
\usepackage{mathrsfs}
\usepackage{upgreek}
\usepackage{titlesec}
\begin{document}

\newcommand{\figref}[1]{\figurename~\ref{#1}}

\title{Universal Quantum Control in Zero-field Nuclear Magnetic Resonance}

\author{Ji Bian$^{1}$, Min Jiang$^{1}$, Jiangyu Cui$^{1}$, Xiaomei Liu$^{1}$, Botao Chen$^{1}$, Yunlan Ji$^{1}$, Bo Zhang$^{1}$, John Blanchard$^{3}$}
\author{Xinhua Peng$^{1,2}$}\email{xhpeng@ustc.edu.cn}
\author{Jiangfeng Du$^{1,2}$}

\affiliation{$^1$Hefei National Laboratory for Physical Sciences at Microscale and Department of Modern Physics, University of Science and Technology of China, Hefei, Anhui 230026, China}
\affiliation{$^{2}$Synergetic Innovation Center of Quantum Information $\&$ Quantum Physics, University of Science and Technology of China, Hefei, Anhui 230026, China}

\affiliation{$^{3}$Helmholtz-Institut Mainz, 55099 Mainz, Germany}


\date{\today}

\begin{abstract}
This paper describes a general method for manipulation of nuclear spins in zero magnetic field. In the absence of magnetic fields, the spins lose the individual information on chemical shifts and inequivalent spins can only be distinguished by nuclear gyromagnetic ratios and spin-spin couplings.
For spin-1/2 nuclei with different gyromagnetic ratios (i.e., different species) in zero magnetic field, we describe the scheme to realize a set of universal quantum logic gates, e.g., arbitrary single-qubit gates and two-qubit controlled-NOT gate. This method allows for universal quantum control in systems which might provide promising applications in materials science, chemistry, biology, quantum information processing and fundamental physics.
\end{abstract}

\maketitle

\titleformat{\section}{\centering\small\bfseries\uppercase}{\thesection}{1em}{}
\section{Introduction}

Zero-field NMR has attracted attention as a tool for chemical analysis \cite{jo,mp,jb,tt,nzf,zulfjb}, not limited by the disadvantages of superconducting magnets typically used in traditional high field NMR. In zero-field NMR, the Zeeman interaction is negligible which provides a natural regime for the measurement of local spin-spin interactions. This is ``the inverse coupling" regime to that in conventional high-field NMR which allows one to measure some complementary information that can not be measured in the high-field case. Zero-field NMR features high absolute field homogeneity and the absence of certain relaxation pathways such as chemical shift anisotropy or susceptibility-induced gradients, yielding narrow resonance lines and accurate determination of coupling parameters \cite{zulfjb,pines}. Very recently, long-lived spin-singlet states (spin-singlet lifetimes as long as 37 seconds) were observed in heteronuclear spin pairs in zero magnetic field \cite{me}, where the lifetime of the singlet-triplet coherence, T$_2$, actually exceeds the lifetime of the triplet-state dipole moment, T$_1$. Further, elimination of expensive cryogenically cooled superconducting magnets enables NMR devices that are portable, affordable, and energy-efficient.

In the absence of an external strong magnetic field, nuclear spin polarization can be prepared through techniques such as parahydrogen-induced polarization \cite{pip,pip2}, dynamic nuclear polarization \cite{dnp,dnp1,dnp2,dnp3}, quantum-rotor induced polarization \cite{qrip,qrip1} or spin-exchange optical pumping \cite{seop,seop1}; encoding can be accomplished through the J-coupling and dipole-dipole coupling between spins; and spin resonance signals can be detected using atomic magnetometers \cite{ik,ik1}, nitrogen-vacancy centers in diamond \cite{nv,nv1}, or superconducting quantum interference devices (SQUIDs) \cite{squids,squids1}. The spins in zero-field NMR can be manipulated by applying pulsed DC fields along three directions ($x$, $y$ and $z$). Unlike high-field NMR, where spin dynamics and control problems are well studied, studies on these topics in zero-field where the spin dynamics and control methods are different from that in high field, are just beginning \cite{m1,m2}.

In this paper, we consider the topic of quantum control in zero-field NMR \cite{unpublished}. While pioneering works have shown that performing arbitrary rotations in zero-field is not a solved problem and generally speaking the control of multiple spin species is significantly restricted \cite{m3,m4}, we show here a way of implementing a set of universal quantum logic gates, i.e., arbitrary single-qubit rotations and two-qubit controlled-NOT gate \cite{qcqi} by using the information on nuclear gyromagnetic ratios and spin-spin couplings. Such a set of gates is sufficient to realize universal control on nuclear spins in zero field.
The controllability in such systems might provide promising applications in materials science, chemistry, biology, quantum information processing and fundamental physics.

\section{Nuclear spin systems in zero magnetic field} \label{sec_sys}

A liquid-state $n$ spin-1/2 system in zero magnetic field can be described by the Hamiltonian ($\hbar$=1):

\begin{equation}
\label{H0}
H_0  = \sum\limits_{i < j, = 1}^n {{2 \pi J_{ij}}}  \mathbf{I}_i \cdot \mathbf{I}_j,
\end{equation}
where ${J_{ij}}$ is the scalar coupling (or J-coupling) constant (in Hz) between the $i$th and $j$th spins and $\mathbf{I}_i = (I_{ix}, I_{iy}, I_{iz})^T$ is the spin angular momentum operator of the $i$th spin:

\begin{equation}
\begin{array}{*{20}{l}}
{{I_x} = \left( {\begin{array}{*{20}{c}}
0&\frac{1}{2}\\
\frac{1}{2}&0
\end{array}} \right)},
{{I_y} = \left( {\begin{array}{*{20}{c}}
0&{ - \frac{i}{2}}\\
\frac{i}{2} &0
\end{array}} \right)},
{{I_z} = \left( {\begin{array}{*{20}{c}}
\frac{1}{2}&0\\
0&{ - \frac{1}{2}}
\end{array}} \right)}.
\end{array}
\end{equation}
We can apply a DC magnetic field $\mathbf{B}$ to such a system:
\begin{equation}
\label{Hdc}
{H_{DC}} (\mathbf{B}) =  -\sum\limits_{i = 1}^n {\gamma _i} \mathbf{B}  \cdot \mathbf{I}_i,
\end{equation}
where $\mathbf{B} = (B_x, B_y, B_z)$ and ${\gamma _i}$ denotes the gyromagnetic ratio of the $i$th spin. DC magnetic field pulses are simultaneously exerted on all the spins, but the effect is dependent on the different gyromagnetic ratios ${\gamma _i}$.

The controllability of such a system is determined by the property of the network of nuclear spins, as studied by Albertini and D'Alessandro \cite{gc}. Taking the spin network as a graph whose nodes represent the spins and whose edges represent the interactions between the two corresponding spins (i.e., there exists an edge between node $i$ and $j$ when $J_{ij} \ne 0$), they relate Lie algebra structure to the properties of a graph.
For networks with different gyromagnetic ratios, the necessary and sufficient condition of controllability is
that the associated graph is connected, which implies that the spin system is completely controllable, i.e., it is possible to realize any unitary element in $SU(2^n)$ for a $n$ spin-1/2 nuclear spin system \cite{gc}. The complete controllability also has significant practical implications, e.g., in quantum information processing, and it is directly related to the question of universality of a quantum computer \cite{qc1,qc2,qc3,qcqi}.

A practical way to achieve universal control with physical operations is to realize a set of universal logic gates, e.g., arbitrary single-qubit gates and two-qubit controlled-NOT gates \cite{qcqi}. The complete controllability tells us that a set of universal logic gates can in principle be implemented in such systems. 
The question is then how to achieve this using the internal Hamiltonian $H_0$ (Eq. \eqref{H0}) and the external Hamiltonian $H_{DC}$ (i.e., DC pulses in Eq. \eqref{Hdc}) in zero-field NMR systems? In the following sections, we answer this question and describe the method to realize a set of universal logic gates consisting of arbitrary single-qubit gates and two-qubit controlled-NOT gate \cite{qcqi}, where the qualities for the operations are evaluated by the gate fidelity \cite{qcqi} defined by
\begin{equation}
F = |\mbox{Tr}(U_{ideal}^{\dag}  U)|/2^n.
\label{fide}
\end{equation}
This describes the accuracy of a realized unitary operation $U$ with respect to the ideal one $U_{ideal}$, and Tr denotes a trace operation.

\section{Arbitrary single-qubit gates}

An arbitrary single-qubit gate on spin $i$ is
\begin{equation}
\label{single_gate}
U_{\mathbf{n}}^ i(\theta) = e^{-i \mathbf{n} \cdot \mathbf{I}_{i}  \theta},
\end{equation}
where $\mathbf{n}$ is the unit vector and $\theta$ is the angle of the rotation.
The available external control is the DC pulse with duration $t$ along any axis with the unit vector $\mathbf{n}$
\begin{equation}
\label{Udc}
U_{DC}({\mathbf{B_n}}) =  e^{-i H_{DC}({\mathbf{B_n}}) t},
\end{equation}
where $H_{DC}({\mathbf{B_n}})$ is given by Eq. \eqref{Hdc} with ${\mathbf{B_n}} = B\mathbf{n}$. Although the spins can not be individually addressed in zero magnetic field, their different gyromagnetic ratios allow one to effectively manipulate them individually. For example, in a two-spin system (say spins $1$ and $2$) \cite{chu}, when
\begin{equation}
\label{cond2}
\frac{\gamma_1}{\gamma_2} = \frac{2m_1+1}{2m_2},
\end{equation}
where $m_{1,2}$ are integers ($ m_2 \neq 0$), one can realize a local $\pi$ pulse on either spin $i$ or spin $j$, e.g., the $^{13}$C ($\gamma_C = 67.262\times10^6 rad \cdot s^{-1} \cdot T^{-1} $)  and $^{1}$H  ($\gamma_H = 267.513\times10^6 rad \cdot s^{-1} \cdot T^{-1} $)  system with $\gamma_C / \gamma_H \approx 1/4$. For any two-spin system with $\gamma_1 \ne \gamma_2$, one can always find the integers $m_1$ and $m_2$ to approximate Eq. \eqref{cond2}. Therefore, an arbitrary single-qubit gate on spin $1$ can be realized by
$$
U_{\mathbf{n}}^ 1(\theta)  =  {U}_{\mathbf{n}_{\perp}} ^{2} (\pi) e^{-i H_{DC}(-{\mathbf{B_n}}) t/2} {U}_{\mathbf{n}_{\perp}} ^{2 \dagger } (\pi) e^{-i H_{DC}(-{\mathbf{B_n}}) t/2} ,
$$
where $\theta = \gamma_1 B t$, $\mathbf{n} \cdot \mathbf{n}_{\perp} = 0$ and ${U}_{\mathbf{n}_{\perp}} ^{2 \dagger }(\pi) = {U}_{\mathbf{n}_{\perp}} ^{2} (-\pi)$ . In this sequence, the phases accumulated by spin $2$ in the two halves of the rotation cancel out, while the phases by spin $1$ are summed to the angle $\theta$. Here we assume that the DC magnetic field $|B| \gg |2 \pi J_{12}/\gamma_1|$ such that we can neglect the effect of J-couplings during the DC pulses. For instance, an arbitrary rotation along the $x$ axis on spin $1$  can be realized as follows
\begin{eqnarray}
{U}_{x}^ 1(\theta) & = & e^{-i I_{1x}  \theta} = \mathcal{U}_{z} ^2 (\pi) e^{-i H_{x} t/2}  \mathcal{U}_{z} ^{2 \dagger }(\pi) e^{-i H_{x} t/2}   \nonumber \\
& = & e^{-i \gamma_1 B_x  I_{1x}   t}   \nonumber
\end{eqnarray}
with $H_x = B_x (\gamma_1 I_{1x} + \gamma_2 I_{2x})$ by using $R e^{-iHt} R^{\dagger} = e^{-iR H R^{\dagger} t} $ with a unitary operator  $R$ and its conjugation $R^{\dagger}$, $\mathcal{U}_{z} ^{2 }(\pi)  I_{2x} \mathcal{U}_{z} ^{2 \dagger } (\pi)  = - I_{2x}$ and $\mathcal{U}_{z} ^{2 }(\pi)  I_{1x} \mathcal{U}_{z} ^{2 \dagger } (\pi)  = I_{1x}$.

Similarly, the realization of any single-qubit gate can be generalized to multi-spin systems:
\begin{equation}
\label{sin}
U_{\mathbf{n}}^1 (\theta) = U_{\mathbf{n}_{\perp}}^{2\sim n} (\pi)  e^{ -i H_{DC}(-{\mathbf{B_n}}) t/2}
   U_{\mathbf{n}_{\perp}}^{2\sim n}{}^{\dagger } (\pi)  e^{ -i H_{DC}(-{\mathbf{B_n}}) t/2},
\end{equation}
where $U_{\mathbf{n}_{\perp}}^{2\sim n} (\pi)=U_{\mathbf{n}_{\perp}}^{2}(\pi)U_{\mathbf{n}_{\perp}}^{3}(\pi)...U_{\mathbf{n}_{\perp}}^{n}(\pi)$.
The key is to implement a $\pi$ rotation on one local spin, e.g., $\mathcal{U}_{\mathbf{n}} ^{j} (\pi) =  e^{-i \mathbf{n} \cdot \mathbf{I}_{j}  \pi}$. Without loss of generality, we consider the case of implementing the target operation $\mathcal{U}_{z} ^{1} (\pi)$, using DC pulse $U_{DC}(B_z)$ in Eq. \eqref{Udc} with $\mathbf{n}$ along the $z$ axes. Intuitively this requires the pulse duration to be such that immediately after the pulse is applied, spin 1  
undergoes a $(2m_1+1)\pi$ rotation around $z$ while spin $j$ ($1<j \leqslant n$) rotates around the same axes by $2m_j \pi$, with integer $m_1$ and $m_j \neq 0$. This is mathematically equivalent to find a pulse duration $t$ satisfying $B_z \gamma_1 t=(2m_1+1)\pi$ and $B_z \gamma_j t=2m_j\pi$,
which has an exact solution if and only if:
\begin{equation}
\label{cond3}
\frac{\gamma_1}{\gamma_j} = \frac{2m_1+1}{2m_j},
\end{equation} 
which is a generalization of $\eqref{cond2}$. A high fidelity $\pi$ pulse can be realized by choosing appropriate m's. As mentioned above, a $\pi$ pulse on $^{13}$C in a $^{13}$C-$^{1}$H system is approximated by choosing $m_1=0,m_2=2$. Take $B_z=9 G$ (the value is in the reasonable range of experimental parameters) and take $t=(2m_1+1)\pi/(B_z\gamma_1)$ to be the pulse duration, one gets 
$$
t = \frac{ \pi}{\gamma_{C} B_z}  = 5.2\times10^{-5}s ,
$$ 
and the gate fidelity is about 0.9994 (gate fidelity is determined by numerical simulation. Alternatively, by \eqref{F1} in the following).

A higher fidelity is achieved by a better approximation of $\eqref{cond3}$, and generally results in a longer pulse duration. This can be seen for example in a $^{31}$P ($\gamma_P = 108.291\times10^6 rad \cdot s^{-1} \cdot T^{-1}$ ) and $^{1}$H ($\gamma_H = 267.513\times10^6 rad \cdot s^{-1} \cdot T^{-1} $) system, one approximate solution of the $\pi$ pulse on $^{31}$P is $m_1 = 2, m_2 = 6$ utilizing the fact that $\gamma_P /\gamma_H  \approx 5/12$. Take the pulse duration to be
$$
t = \frac{ 5 \pi}{ \gamma_{P} B_z}  = 1.6 \times10^{-4}s ,
$$
with $B_z = 9$ G , the corresponding gate fidelity is about 0.9782.
A higher fidelity can be obtained by a better approximation of Eq. \eqref{cond3}, e.g., $ 17/42 $ is closer to the real value of $\gamma_P /\gamma_H$ than $5/12$, so $m_1 = 8$ and $m_2 = 21$ (utilizing $\gamma_P /\gamma_H  \approx 17/42$) is a better approximation. And indeed it results in a higher gate fidelity ($\sim 0.9998$) with a longer pulse length $t = 5.48\times10^{-4} s$.

Pulse durations in multi-spin systems can also be determined through approximating \eqref{cond3}. Appropriate m's have to be chosen to approximate $n-1$ equations simultaneously in \eqref{cond3} in a $n$-spin system. Alternatively, write out the function of fidelity with respect to pulse duration and choose a duration with high enough fidelity. This is done in the following (omit the effect of J-coupling ): with the target operation $\mathcal{U}_{z} ^{1} (\pi)$ and DC pulse $U_{DC}(B_z)$, the gate fidelity becomes: 
\begin{equation}
\begin{aligned}
\label{F1}
F &=|Tr(\mathcal{U}_{z} ^{1} (\pi)^{\dag}  U_{DC}(B_z))|/2^n\\
  &=\left |  \sin{\frac{B_z \gamma_1 t}{2}} \right | \prod_{j=2}^n{\left| \cos{\frac{B_z \gamma_j t}{2}} \right|},
\end{aligned}
\end{equation}
which is a product of $n$ separated gate fidelities, each defined in a single-spin system. 
\begin{figure}[htbp]
\begin{center}
\includegraphics[scale=0.25]{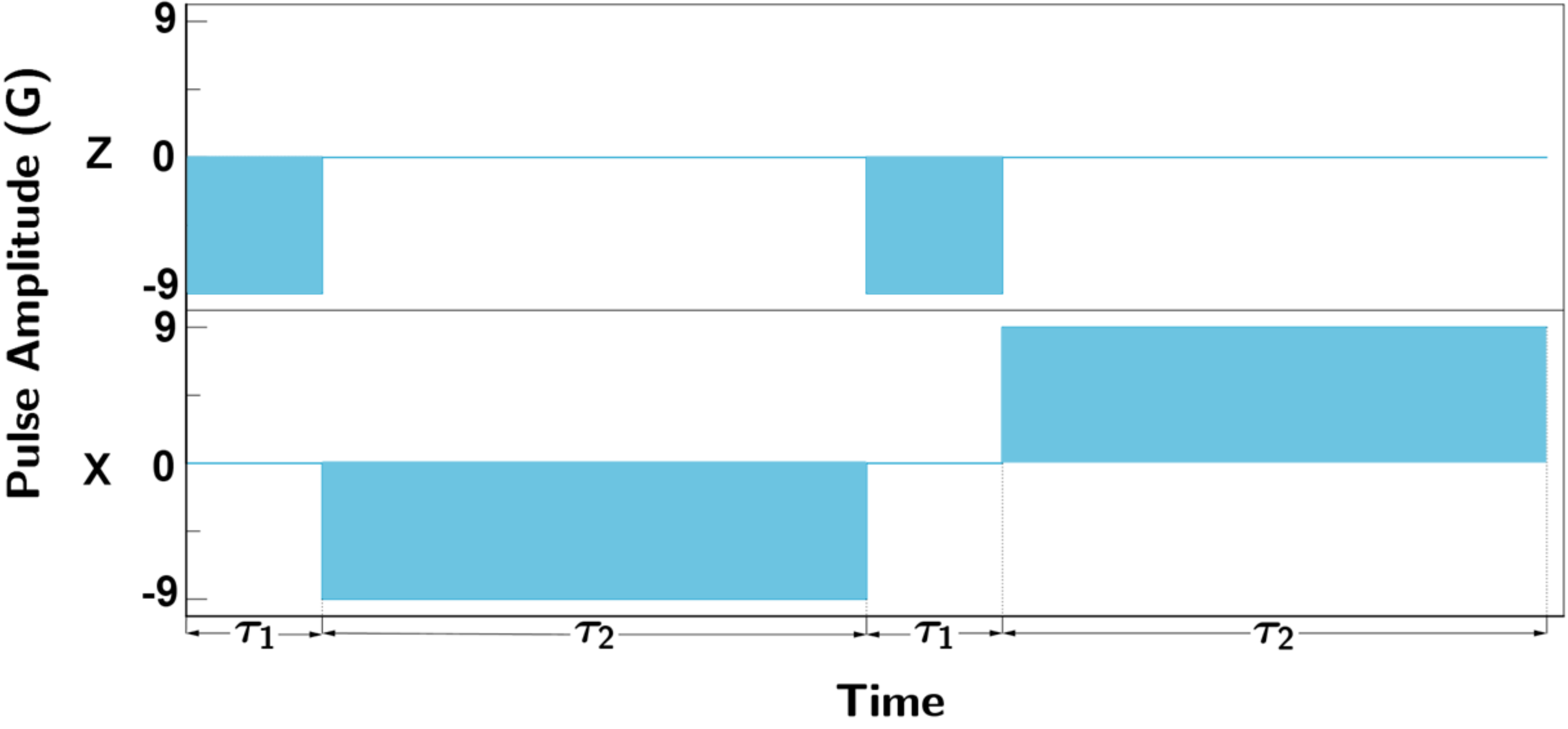}
\caption{Pulse sequence realizing $\pi/2$ gate on $^{13}$C in C-H-F system. $\tau_1 = \frac{\pi}{4\gamma_C B_z} \approx 12.9 \mu $s. $\tau_2 \approx 1761 \mu$s is the time to realize $\pi$ pulse on both $^{13}$C and $^{19}$F simultaneously (leaving H alone), which is found by numerical simulation discussed in the main text.}
\label{xmct}
\end{center}
\end{figure}
\eqref{F1} equals to $1$ if and only if \eqref{cond3} holds, which is in consistent with the above discussion.
As an example, via \eqref{F1}, the approximate solution of the $\pi$ pulse on $^{19}$F in a three-spin system consisting of $^{13}$C, $^{1}$H and $^{19}$F ($\gamma_F = 251.662\times10^6 rad\cdot s^{-1} \cdot T^{-1}$) is $t =2.1\times10^{-4}s $ and $F=0.9932 $ for $B_z = 9$ G.  

Generally, for systems composed of a larger number of spins, it requires a longer time to implement a $\pi$ pulse (perhaps with a lower fidelity). However, in a small spin system ($\sim 3-5$ spins), it is feasible to find a reasonable solution with a high enough fidelity to achieve the local rotation by this method. For instance, in the C-H-F system, the $\pi /2$ pulse ${U}_{\mathbf{n}}^ {C}(\pi /2)$ is constructed with the gate fidelity around 0.9998 via Eq. \eqref{sin}, where the $\pi$ pulses are implemented as above. The whole sequence is illustrated in \figref{xmct}. By a similar procedure, local rotations on two or more different spins can also in principle be achieved.

Very recently, other methods have been found to achieve control with spin-species selectivity \cite{m1} or transition selectivity \cite{m2} in zero-field NMR. For example, the high-field selectivity in zero-field NMR is used by temporarily applying a magnetic field on the sample, allowing one to apply AC pulses that individually address different spin species, like that in high-field NMR \cite{m1}. In principle, this method is feasible for implementation of arbitrary single-qubit gates if all gyromagnetic ratios are different, as the operators are almost the same as those in high-field NMR quantum information processing. Furthermore, transition-selective pulses have been demonstrated in zero-field NMR \cite{m2} which can also be implemented by the set of universal logic gates presented in this paper.

\section{Two-qubit controlled-NOT gate}

In order to to achieve universal control on a multi-spin system, one still needs a two-qubit gate, e.g. the controlled-NOT gate between spin $i$ and spin $j$. Its matrix form in $I_z$ basis $\{ |0 \rangle_i | 0 \rangle_j ,  |0 \rangle_i | 1 \rangle_j,  |1 \rangle_i | 0 \rangle_j,  |1 \rangle_i | 1 \rangle_j\}$ with $I_z |0 \rangle = \frac{1}{2}|0 \rangle $ and  $I_z |1 \rangle = - \frac{1}{2}|1 \rangle $ reads:
\begin{equation}
\label{ocnot}
CNOT_{ij} = \left( {\begin{array}{*{20}{c}}
1&0&0&0\\
0&1&0&0\\
0&0&0&1\\
0&0&1&0
\end{array}} \right),
\end{equation}
where spin $i$ (the high bit) is the control spin, and spin $j$ (the low bit) is the target spin. This operation flips spin $j$ (target spin) when spin $i$ (control spin) is in the state $|1\rangle$ and doing nothing when spin $i$ is in the state $|0\rangle$. This operation can be further decomposed into \cite{chu}
\begin{equation}
\label{CNOT}
\text{CNOT}_{ij} =\sqrt{i} U_z^i (\frac{\pi}{2})  U_z^j{}^{\dagger } (\frac{\pi}{2})  U_x^j (\frac{\pi}{2})
   U^{(i,j)}_{\text{zz}} (\frac{\pi}{2})  U_y^j (\frac{\pi}{2});
\end{equation}
in which
\begin{equation}
 U^{(i,j)}_{\text{zz}} (\theta)=e^{-iH^{(i,j)}_0t}U_z^j(\pi)e^{-iH^{(i,j)}_0t}U_z^j{}^{\dagger }(\pi),
 \label{Uzz}
 \end{equation}
 where $H^{(i,j)}_0= {{2 \pi J_{ij}}}  \mathbf{I}_i \cdot \mathbf{I}_j$ and $\theta \equiv 2 \pi J_{ij}t$ for $J_{ij}>0$. Arbitrary single-qubit gate $U_{\alpha}^{i}$ or $U_{\alpha}^{j} (\theta) (\alpha = x, y \mbox{ or } z)$ is realized by the method in Sec. III. If $J_{ij}<0$,  $ \text{CNOT}_{ij}$ is realized by $\text{CNOT}_{ij}^{\dagger}$ as $ \text{CNOT}_{ij} = \text{CNOT}_{ij}^{\dagger}$ and the free evolution under $H^{(i,j)}_0$ is a conjugate to one with the case of $J_{ij} >0$.

For spin systems with $n$ spins ($n>2$), the main barrier to implementing the controlled-NOT gate is the implementation of $U^{(i,j)}_{\text{zz}} (\theta)$ in a large coupled spin network, where only the coupling $H^{(i,j)}_0$ is active.
To achieve this, one needs to turn off the undesired couplings, as achieved by refocusing schemes \cite{rf} in high-field NMR. This is, however, somewhat more complicated in zero-field NMR.

\begin{figure*}[hbp]
\begin{center}
\includegraphics[scale=0.5]{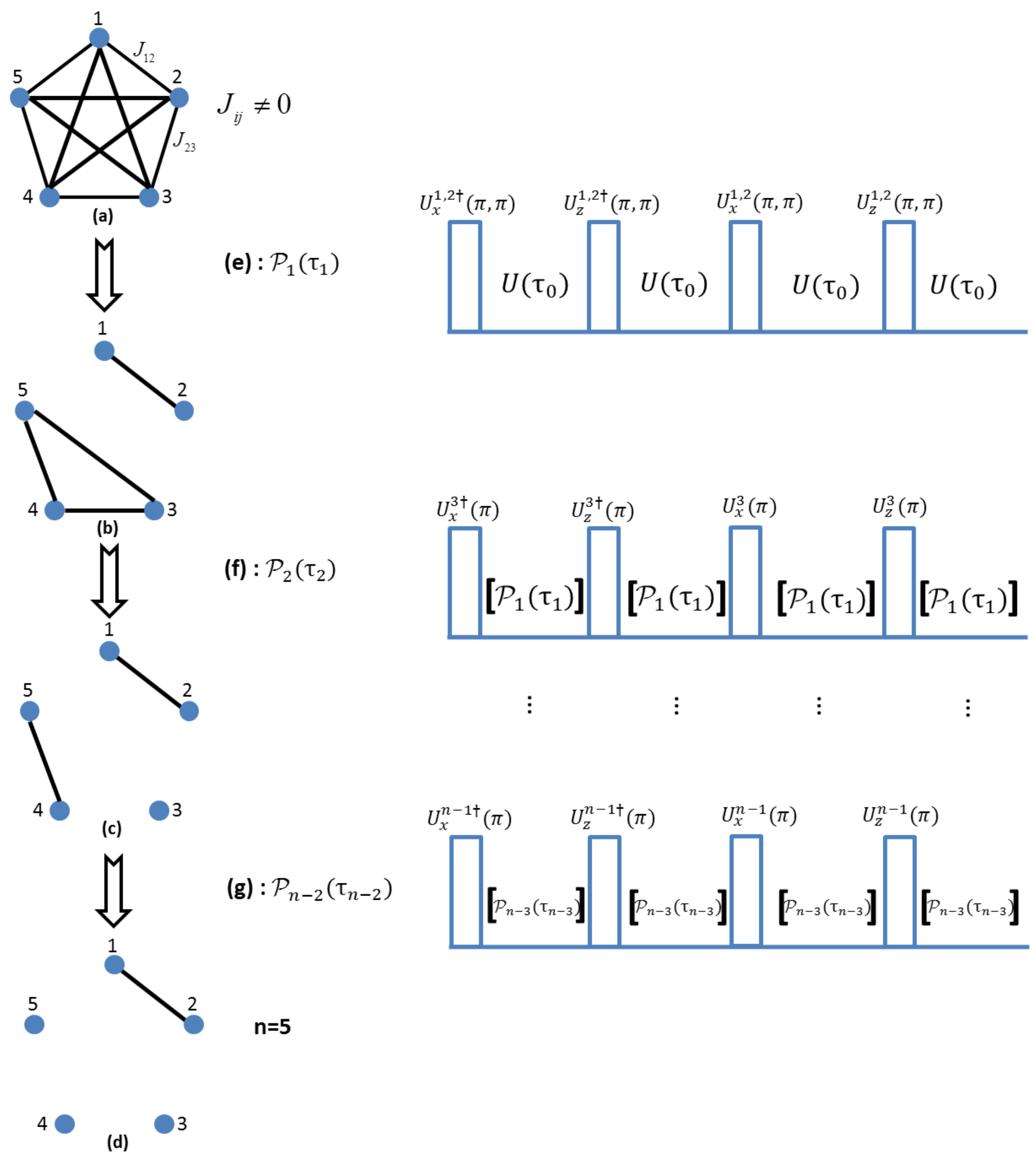}
\caption{Multiple pulse sequence for implementing the zero-order average Hamiltonian $\bar{\mathcal{H}}^{(0)} = H^{(1,2)}_0 $ in a complex $n$-spin network: (a) - (d) for the diagram of the spin network and (e) - (g) for the corresponding concatenated pulse sequence.}
\label{cnotzh}
\end{center}
\end{figure*}

Consider a complex spin network where all spin pairs are coupled, e.g., an example shown in \figref{cnotzh} (a). Let us first analyze a basic pulse sequence shown in \figref{cnotzh} (e):
\begin{equation}
\begin{aligned}
\mathcal{U} &=  U(\tau_0)U_z^{1... n}(\theta_{1},...\theta_{n})U(\tau_0)U_z^{1...n} {}^\dagger(\theta_{1},...\theta_{n})\\
&\times U_y^{1...n}(\theta_{1},...\theta_{n})U(\tau_0)U_y^{1...n} {}^\dagger(\theta_{1},...\theta_{n})\\
&\times U_x^{1...n}(\theta_{1},...\theta_{n})U(\tau_0)U_x^{1...n} {}^\dagger(\theta_{1},...\theta_{n})\\
\end{aligned}
\label{ul}
\end{equation}
with $U(\tau_0) = e^{-iH_0\tau_0} $ and $U_{\alpha}^{1,2...n} (\theta_{1},\theta_{2}...\theta_{n})=U_{\alpha}^{1}(\theta_1)U_{\alpha}^{2}(\theta_2)...U_{\alpha}^{n}(\theta_n), ({\alpha}=x,y,z)$, ($\theta_{m}=0$ if $m$ does not appear on the upper index of $U_{\alpha}^{1,...}(\theta_{1},...)$).
By average Hamiltonian theory \cite{st}, one gets the zero-order approximation for $\tau_0 \to 0$:
\begin{equation}
\mathcal{U} \approx  e^{-i\tau_0 [H_0+\sum\limits_{{\alpha}=x,y,z} {U_{\alpha}^{1...n}(\theta_{1}...\theta_{n})H_0U_{\alpha}^{1...n}{}^\dagger(\theta_{1}...\theta_{n})]}}.
\label{app}
\end{equation}
Here
\begin{small}
\begin{equation}
\begin{aligned}
& U_{\alpha}^{1...n} (\theta_{1}...\theta_{n}) H_0 U_{\alpha}^{1...n} {}^\dagger (\theta_{1}...\theta_{n}) \\
&=\sum\limits_{i<j, =1}^n U_{\alpha}^{1...n} (\theta_{1}...\theta_{n}){H_0^{(i,j)}}U_{\alpha}^{1...n} {}^\dagger (\theta_{1}...\theta_{n})\\
&=\sum\limits_{i<j, =1}^n {2\pi J_{ij}[I_{i{\alpha}}I_{j{\alpha}}+I_{i{\beta}}I_{j{\beta}}\cos(\theta_i-\theta_j)+I_{i\gamma}I_{j\gamma}\cos(\theta_i-\theta_j)]}\\
&+\sum\limits_{i<j, =1}^n{2\pi J_{ij}[I_{i\gamma}I_{j\beta}\sin(\theta_i-\theta_j)+I_{i\beta}I_{j\gamma}\sin(\theta_j-\theta_i)]},
\end{aligned}
\label{exp}
\end{equation}
\end{small}
where $\{\alpha,\beta,\gamma\}$ is cyclic permutation of $ \{x,y,z\}$.
Relations like $e^{-i\theta I_{i\alpha}}I_{i\alpha}e^{i\theta I_{i\alpha}} = I_{i\alpha}$, and $e^{-i\theta I_{i\alpha}}I_{i\beta}e^{i\theta I_{i\alpha}}=I_{i\beta}\cos \theta+I_{\i\gamma}\sin \theta$ are used.  Hence one gets
\begin{eqnarray}
&& H_0^{(i,j)}+\sum_{\alpha = x,y,z}U_{\alpha}^{1...n} (\theta_{1}...\theta_{n}) H_0^{(i,j)} U_{\alpha}^{1...n} {}^\dagger (\theta_{1}...\theta_{n}) \nonumber \\
&=& \left\{
\begin{array}{ll}
 0, &    \theta_i-\theta_j=( 2r_{ij} -1) \pi,  \\
 4H^{(i,j)}_0,  &    \theta_i-\theta_j=2r_{ij}\pi,
\end{array} \right.
\label{conth}
\end{eqnarray}
where $r_{ij}$ is any integer. This property shows that one can turn on or turn off the coupling $H^{(i,j)}_0$ by choosing the rotation angles $\theta_i$ and $\theta_j$. The simplest choices of $\theta_i$ and $\theta_j$ are the integer multiples of $\pi$. For example, by setting
$$
\theta_1=\theta_2 = \pi,  \theta_j = 0 \mbox{ for } j= 3,4... .
$$
Eq. \eqref{ul} is rewritten as:
\begin{equation}
\begin{aligned}
\mathcal{U} =&U(\tau_0)U_{z}^{1,2} (\pi,\pi)U(\tau_0)U_{z}^{1,2} {}^\dagger(\pi,\pi)U_{y}^{1,2} (\pi,\pi)\\
&\times U(\tau_0)U_{y}^{1,2} {}^\dagger (\pi,\pi)
U_{x}^{1,2} (\pi,\pi)U(\tau_0)U_{x}^{1,2} {}^\dagger(\pi,\pi)\\
=&U(\tau_0)U_{z}^{1,2} (\pi,\pi)U(\tau_0)U_{x}^{1,2} (\pi,\pi)\\
&\times U(\tau_0)
U_{z}^{1,2} {}^\dagger (\pi,\pi)U(\tau_0)U_{x}^{1,2} {}^\dagger(\pi,\pi),
\end{aligned}
\end{equation}
up to a normalized phase factor, as shown in \figref{cnotzh} (e).  Thus one gets from Eq. \eqref{conth}, the average Hamiltonian during the pulse sequence is
\begin{eqnarray}
 \bar{\mathcal{H}}^{(0)} & = & H_0+\sum_{\alpha = x,y,z}U_{\alpha}^{1,2}(\pi,\pi)H_0U_{\alpha}^{1,2}{}^\dagger(\pi,\pi) \nonumber \\
&=&4( H_0^{(1,2)}+\sum\limits_{i<j, =3}^n H_0^{(i,j)}).
\label{le11}
\end{eqnarray}

After the sequence, the spin network shown in \figref{cnotzh} (a) is decoupled into two uncoupled subsystems: the pair of spin 1 and 2, and the rest network consisting of all the other spins $3,...,n$, as shown in \figref{cnotzh} (b). However, the implementation of a CNOT$_{ij}$ gate in an $n$-qubit system requires keeping only $2\pi J_{ij}\mathbf{I}_i \cdot \mathbf{I}_j$ while turning off all of the other couplings. This can be achieved via a concatenated scheme by \emph{recursively} building on the base sequence $[\cdot ]Z_k[\cdot
]X_k[\cdot ]Z_k^{\dagger}[\cdot ]X_k^{\dagger}$, as shown in \figref{cnotzh}.
Here $X_k \equiv U_x^{i,j}(\pi, \pi) (k=1) \mbox{ or } U_x^{k+1}(\pi) (k>1)$ and  $Z_k \equiv U_z^{i,j} (\pi, \pi) (k=1) \mbox{ or } U_z^{k+1}(\pi) (k>1)$. The sequence is initialized as
\begin{equation}
\mathcal{P}_{0}(\tau_0) =U(\tau_0)
\end{equation}%
and higher levels are generated via the rule
\begin{small}
\begin{eqnarray}
\mathcal{P}_{k+1}(\tau_{k+1}) = [\mathcal{P}_{k}(\tau
_{k})]Z_k[\mathcal{P}_{k}(\tau _{k})]X_k[\mathcal{P}_{k}(\tau _{k})]Z_k^{\dagger}[\mathcal{P}_{k}(\tau _{k})]X_k^{\dagger},  \nonumber
\label{eq:CDD-def}
\end{eqnarray}
\end{small}
where $\tau _{k}=4^{k}\tau _{0}$. By setting $\theta_1= \theta_2=\pi$ and $\theta_m=0$ for $m>2$ in the first-level  $\mathcal{P}_{1}$, an $n$-spin coupled system is divided into two subsystems: $1+2$ and $3+4+...+n$. Spin 3 is further decoupled from the subsystem $3+4+...+n$ and keeps the $1+2$ subsystem unchanged in the second-level $\mathcal{P}_{2}$ with $\theta_3=\pi$ and $\theta_m=0$ for $m \neq 3$. The $(n-2)^{th}$-level procedure is required until all the spins in the subsystem $3+4+...+n$ are decoupled, and the coupling between spin 1 and 2 is kept. The procedure is shown schematically in \figref{cnotzh}. Thus $U^{(1,2)}_{\text{zz}}(\theta)= e^{-i H^{(1,2)}_0t}U_z^{1}(\pi)  e^{-i H^{(1,2)}_0 t}U_z^{1}{}^{\dagger }(\pi) $ with $\theta = 4 \pi J_{12} t$. In order to implement the CNOT$_{12}$ gate, the total time under $H^{(1,2)}_0$ is $T = 4 \tau _{n-3}=4^{n-2}\tau _{0} = \frac{1}{4J_{12}}$.

For a three-spin system, CNOT$_{12}$ is realized by the first-level sequence with $\tau_0=1/(16J_{12})$:
\begin{eqnarray}
\mathcal{P}_1: &  \theta_1 = \theta_2= \pi, & \theta_3 = 0.   \nonumber
\end{eqnarray}
The pulse sequence for realizing CNOT$_{CH}$ is numerically simulated for the $^{13}$C-$^{1}$H-$^{19}$F system (diethyl fluoromalonate) \cite{pls} with J-coupling constants: $J_{12} = 160.7$ Hz, $J_{13} = -194.4$ Hz, $J_{23} = 47.6$ Hz.
The gate fidelity is about 0.9993 if the J coupling is neglected during the evolutions and all single-qubit gates required are assumed to be perfect. If single-qubit gates are achieved via the method discussed in section III, the gate fidelity is about 0.9927.

The procedure can be slightly modified to simultaneously realize several non-connected CNOT$_{ij}$ operations. For example, the zero-order average Hamiltonian $ \bar{\mathcal{H}}^{(0)} = H^{(1,2)}_0 + H^{(3,4)}_0$ can be generated by setting
$\theta_3= \theta_4 = \pi,  \theta_j = 0 \mbox{ for } j \neq 3 \mbox{ or } 4$ in the second-level sequence $\mathcal{P}_{2}$, while maintaining the rest of the above procedure unchanged ($J_{12},J_{34}$ is unchanged while the rest couplings are turned off).
$U_{\text{zz}}^{(1,2)} (\theta_1)$ and $U_{\text{zz}}^{(3,4)} (\theta_2)$ can be simultaneously implemented by
\begin{equation}
\label{Uzz1234}
\begin{aligned}
U_{\text{zz}}^{(1,2)} (\theta_1) U_{\text{zz}}^{(3,4)} (\theta_2) = &  e^{-i(H^{(1,2)}_0+H^{(3,4)}_0)t}U_{\text{z}}^{2,4}(\pi,\pi)\\
&  \times e^{-i(H^{(1,2)}_0+H^{(3,4)}_0)t}U_{\text{z}}^{2,4}{}^{\dagger }(\pi,\pi)
\end{aligned}
 \end{equation}
with $\theta_1 = 2 \pi J_{12} t$ and $\theta_2 = 2 \pi J_{34} t$. When $J_{12} = J_{34}$,
$U_{\text{zz}}^{(1,2)} (\frac{\pi}{2})$ and $U_{\text{zz}}^{(3,4)} (\frac{\pi}{2})$ can be simultaneously, directly implemented by Eq. \eqref{Uzz1234}. When $J_{12}\neq J_{34}$, e.g., $J_{12}< J_{34}$,
\begin{equation}
\begin{aligned}
U_{\text{zz}}^{(1,2)} (\frac{\pi}{2}) U_{\text{zz}}^{(3,4)} (\frac{\pi}{2})  = & e^{-i H^{(1,2)} t_2} e^{-i(H^{(1,2)}_0+H^{(3,4)}_0)t_1}\\
& \times U_{\text{z}}^{2,4}(\pi,\pi) e^{-i H^{(1,2)} t_2} \\
&  \times e^{-i(H^{(1,2)}_0+H^{(3,4)}_0)t_1} U_{\text{z}}^{2,4}{}^{\dagger }(\pi,\pi)  \nonumber
\end{aligned}
 \end{equation}
where $t_1 = \frac{1}{4 J_{34}}$ and $t_2 = \frac{1}{4 J_{12}} - \frac{1}{4 J_{34}}$. Therefore, CNOT$_{12}$ and CNOT$_{34}$ can be simultaneously implemented via:
\begin{equation}
\begin{aligned}
\text{CNOT}_{12;34} =&i U_z^{1,3} (\frac{\pi}{2},\frac{\pi}{2})  U_z^{2,4}{}^{\dagger } (\frac{\pi}{2},\frac{\pi}{2})  U_x^{2,4} (\frac{\pi}{2},\frac{\pi}{2})\\ \nonumber
   & \times U_{\text{zz}}^{(1,2)} (\frac{\pi}{2}) U_{\text{zz}}^{(3,4)} (\frac{\pi}{2})U_y^{2,4} (\frac{\pi}{2},\frac{\pi}{2}).
  \end{aligned}
 \end{equation}
Without loss of generality, $J_{ij} > 0 $ is assumed.

\section{Conclusion}

In summary, we have discussed the topic of universal control in zero-field NMR. Unlike the case in high-field NMR where nuclear spins can be individually addressed by different frequencies of RF irradiation, here nuclear spins are distinguishable by different gyromagnetic ratios and/or J-coupling constants. A general method is developed to design the pulse sequences for implementing a set of universal logic gates, i.e., arbitrary single-qubit gates and a two-qubit controlled-NOT gate for nuclear spins in zero magnetic field where all spins have the different gyromagnetic ratios. This provides an operational method to achieve the universal control for such systems. This method is experimentally feasible for some small real spin systems, such as formic acid \cite{me}, diethyl fluoromalonate \cite{pls} , acetonitrile \cite{nzf} and so on. While the method can in principle be applied to some large spin systems , the exponential scaling of the free evolution time and the number of $\pi$ pulses, together with the increasing of each $\pi$ pulse duration, will always limit its practicability to within systems with small number of qubit. 
 
Moreover, attention should be paid to some simplifications with neglecting the effect of J-coupling, the relaxation and magnetic field inhomogeneity in the calculation of the gate fidelity. Like in high field, we can combine further this current method with the methods of self-refocusing shaped pulses \cite{sr,gf}, composite pulses \cite{cp} and numerical optimization \cite{GRAPE} and so on. The numerical method is currently underway as our next work and will be described elsewhere. We expect the study of universal control in zero-field NMR will offer promising applications in materials science, chemistry, biology, quantum information processing and fundamental physics.

\mbox{}\\
\section{Acknowledgements}

We thank Prof. Dmitry Budker for helpful discussions and comments. This work is supported by National Key Basic Research Program of China (2013CB921800 and 2014CB848700), the National Science Fund for Distinguished Young Scholars (Grants No. 11425523), the National Natural Science Foundation of China (Grants No. 11375167 and No. 11227901), the Strategic Priority Research Program (B) of the CAS (Grant No. XDB01030400). Key Research Program of Frontier Sciences of the CAS (Grant No. QYZDY-SSW-SLH004).


\end{document}